\documentclass[twocolumn,showpacs,amsmath,amssymb]{revtex4}
\usepackage{graphicx}% Include figure files
\usepackage{dcolumn}% Align table columns on decimal point
\usepackage{bm}% bold math
\usepackage{psfig}% bold math
\newcommand{\kskl}{K_S^0 K_L^0}
\newcommand{\ks}{K_S^0}
\newcommand{\kl}{K_L^0}
\newcommand{\BR}{{\cal B}}
\newcommand{\eff}{\varepsilon}

\newcommand{\psp}{\psi(2S)}
\newcommand{\jpsi}{J/\psi}
\newcommand{\etac}{\eta_{c}}

\newcommand{\EE}{e^+e^-}
\newcommand{\MM}{\mu^+\mu^-}

\newcommand{\piz}{\pi^0}
\newcommand{\pp}{\pi^+\pi^-}
\newcommand{\kk}{K^+K^-}
\newcommand{\ppb}{p\overline{p}}

\newcommand{\kskn}{\overline{K}^*(892)^0 K^0 + c.c.}
\newcommand{\kskc}{\overline{K}^*(892)^- K^+ + c.c.}

\newcommand{\jpsipp}{\pi^+\pi^-J/\psi}

\newcommand{\ra}{\rightarrow}
\newcommand{\jpsito}{J/\psi \rightarrow }

\newcommand{\pspto}{\psi(2S) \rightarrow }

\newcommand{\bfg}{\begin{figure}}
\newcommand{\efg}{\end{figure}}
\newcommand{\bitm}{\begin{itemize}}
\newcommand{\eitm}{\end{itemize}}
\newcommand{\bnum}{\begin{enumerate}}
\newcommand{\enum}{\end{enumerate}}
\newcommand{\btbl}{\begin{table}}
\newcommand{\etbl}{\end{table}}
\newcommand{\btbu}{\begin{tabular}}
\newcommand{\etbu}{\end{tabular}}
\begin{document}

\preprint{Draft-PRD}

\title{\boldmath Improved measurement of the branching ratio of $\jpsito \kskl$}
\author{
J.~Z.~Bai$^1$,        Y.~Ban$^{10}$,         J.~G.~Bian$^1$,
X.~Cai$^{1}$,         J.~F.~Chang$^1$,       H.~F.~Chen$^{16}$,    
H.~S.~Chen$^1$,       H.~X.~Chen$^{1}$,      J.~Chen$^{1}$,        
J.~C.~Chen$^1$
% ...
\\(BES Collaboration)\\ 
$^1$ Institute of High Energy Physics, Beijing 100039, People's Republic of
     China\\
$^2$ China Center of Advanced Science and Technology, Beijing 100080,
     People's Republic of China\\
$^3$ Guangxi Normal University, Guilin 541004, People's Republic of China\\
$^4$ Henan Normal University, Xinxiang 453002, People's Republic of China\\
$^5$ Huazhong Normal University, Wuhan 430079, People's Republic of China\\
$^6$ Hunan University, Changsha 410082, People's Republic of China\\                                                  
$^7$ Liaoning University, Shenyang 110036, People's Republic of China\\
$^{8}$ Nanjing Normal University, Nanjing 210097, People's Republic of China\\
$^{9}$ Nankai University, Tianjin 300071, People's Republic of China\\
$^{10}$ Peking University, Beijing 100871, People's Republic of China\\
$^{11}$ Shandong University, Jinan 250100, People's Republic of China\\
$^{12}$ Shanghai Jiaotong University, Shanghai 200030, 
        People's Republic of China\\
$^{13}$ Sichuan University, Chengdu 610064,
        People's Republic of China\\                                    
$^{14}$ Tsinghua University, Beijing 100084, 
        People's Republic of China\\
$^{15}$ University of Hawaii, Honolulu, Hawaii 96822\\
$^{16}$ University of Science and Technology of China, Hefei 230026,
        People's Republic of China\\
$^{17}$ Wuhan University, Wuhan 430072, People's Republic of China\\
$^{18}$ Zhejiang University, Hangzhou 310028, People's Republic of China
}

\date{\today}
           
\begin{abstract}

The branching ratio of $\jpsito \kskl$ is measured with
improved precision to be 
\( \BR(\jpsito \kskl) = (1.82\pm 0.04\pm 0.13)\times 10^{-4}\)
using $\jpsi$ data collected with the Beijing Spectrometer (BESII)
at the Beijing Electron-Positron Collider.
This result is used to test the perturbative QCD ``12\%'' rule
between $\psp$ and $\jpsi$ decays and to investigate the 
relative phase between the three-gluon and one-photon annihilation 
amplitudes in $\jpsi$ decays.
    
\end{abstract}


\maketitle

\section{Introduction}

\subsection{\boldmath Decays of $\jpsito \kskl$ }

The decays of the $\jpsi$ into light hadronic final states can proceed
via either three-gluon or one-photon annihilations, and it has been
determined that the phases of these amplitudes are nearly orthogonal
in many two-body exclusive decays, such as Vector-Pseudoscalar (VP),
Vector-Vector (VV), Pseudoscalar-Pseudoscalar (PP) and Nucleon
anti-Nucleon
(N$\overline{\hbox{N}}$)~\cite{suzuki,dm2exp,mk3exp,a00,a11,ann}.  For
the PP phase analysis, the $\pp$, $\kk$, and $\kskl$ decay branching
ratios are required~\cite{haber,a00,a11}.  The available $\jpsito
\kskl$ branching ratios come from DMII~\cite{dm2exp} and
MARKIII~\cite{mk3pp}; these measurements have relative errors of about
18\%. Here we report a measurement of the $\kskl$ decay branching
fraction using the $\jpsi$ data sample collected with the Beijing
Spectrometer (BESII) at the Beijing Electron-Positron Collider (BEPC).

Furthermore, there is a prediction of the relation between $\jpsi$
and $\psp$ decay branching ratios to the same hadronic final 
state ($h$)~\cite{applequist,franklin}, that is
      \[ Q_h = \frac{\BR(\pspto h)}{\BR(\jpsito h)}
             =\frac{\BR(\pspto e^+e^-)}{\BR(\jpsito e^+e^-)}
             \approx 12\%. \]
While some channels obey the so called ``12\% rule'',
others violate this rule very badly~\cite{franklin,wangwf}.
Thus it is interesting to test this rule for $\kskl$ decay, which
can only be produced through SU(3) symmetry-breaking, strong
decays of these charmonium states.

\subsection{The experiment}

The data used for this analysis are taken with the
BESII detector at the BEPC storage ring
at a center-of-mass energy corresponding to $M_{\jpsi}$.
The data sample corresponds to a total of 
$57.7(1 \pm 4.7\%)\times 10^6$ $\jpsi$ decays, as determined 
from inclusive 4-prong hadrons~\cite{fangss}.

BES is a conventional solenoidal magnet detector that is
described in detail in Ref.~\cite{bes}; BESII is the upgraded version
of the BES detector~\cite{bes2}. A 12-layer vertex
chamber (VC) surrounding the beam pipe provides trigger
information. A forty-layer main drift chamber (MDC), located
radially outside the VC, provides trajectory and energy loss
($dE/dx$) information for charged tracks over $85\%$ of the
total solid angle.  The momentum resolution is
$\sigma _p/p = 0.017 \sqrt{1+p^2}$ ($p$ in $\hbox{\rm GeV}/c$),
and the $dE/dx$ resolution for hadron tracks is $\sim 8\%$.
An array of 48 scintillation counters surrounding the MDC  measures
the time-of-flight (TOF) of charged tracks with a resolution of
$\sim 200$ ps for hadrons.  Radially outside the TOF system is a 12
radiation length, lead-gas barrel shower counter (BSC).  This
measures the energies
of electrons and photons over $\sim 80\%$ of the total solid
angle with an energy resolution of $\sigma_E/E=22\%/\sqrt{E}$ ($E$
in GeV).  Outside of the solenoidal coil, which
provides a 0.4~Tesla magnetic field over the tracking volume,
is an iron flux return that is instrumented with
three double layers of  counters that
identify muons of momentum greater than 0.5~GeV/$c$.

\subsection{Monte Carlo}
A Monte Carlo simulation is used for the determinations of the mass
resolution and detection efficiency.  This program
(SIMBES), which is Geant3 based, simulates the detector response,
including the interactions of secondary particles with the detector
material.  Reasonable agreement between data and Monte Carlo
simulation has been observed in various channels tested, including
$\EE \ra (\gamma)\EE$, $\EE\ra (\gamma)\MM$, $\jpsito \ppb$ and
$\pspto \jpsipp, \jpsito \ell^+\ell^-$ $(\ell=e,\mu)$.

For the signal channel, $\jpsito \kskl$, the angular distribution of
the $\ks$ or $\kl$ is generated as $\sin^2\theta$, where $\theta$ is
the polar angle in the laboratory system. The $\kl$ is allowed to
decay and to interact with the detector material, and for the $\ks$,
only $\ks \ra \pp$ is generated. For this study, 50,000 events are
generated.  A Monte Carlo sample with 30~M inclusive $\jpsi$ decays
generated with Lundcharm~\cite{lundcharm} is used for background
estimation.

\section{Event selection}

For the decay channel of interest, the candidate events
must satisfy the following selection criteria:
\begin{enumerate}
\item   Two charged tracks
        with net charge zero are required.

\item   Each track should satisfy
        $|\cos\theta|<0.80$, where $\theta$ is the polar angle 
        in the MDC, and have a good helix
        fit so that the error matrix from track fitting is
        available for secondary vertex finding. 

\item To remove backgrounds mainly from $\jpsito \kskn$,
        $E_{\gamma}^{lft}<0.1$~GeV is required, where
        $E_{\gamma}^{lft}$ is the sum of the energies of the photon
        candidates outside a cone about the direction of the $\kl$
        ($\cos\theta<0.95$). A neutral cluster is considered to be a
        photon candidate when the angle between the nearest charged
        track and the cluster in the $xy$ plane is greater than
        $15^{\circ}$, the first energy deposit is in the beginning 6
        radiation lengths, and the angle between the cluster
        development direction in the BSC and the photon emission
        direction in the $xy$ plane is less than $37^{\circ}$.
\end{enumerate}

\begin{figure}[htbp]
\centerline{\hbox{
\psfig{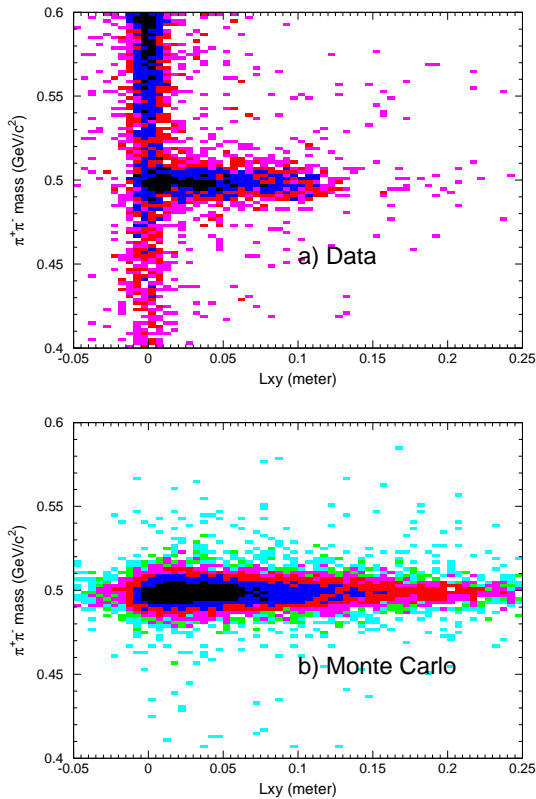}}}
\caption{Scatter plot of $\pp$ invariant mass versus the decay 
length in the transverse plane for events with $\ks$ momentum between
1.45 and 1.50~GeV/$c$ for a) data and b) Monte Carlo simulation.}
\label{mkslxy-j}
\end{figure}

The two tracks are assumed to be $\pi^+$ and $\pi^-$.  To find the
intersection of the two tracks near the interaction point, an
iterative, nonlinear least squares technique is
used~\cite{wangzhe}. The intersection is taken as the $\ks$ vertex,
and the momentum of the $\ks$ is calculated at this point.
Figure~\ref{mkslxy-j} shows a scatter plot of the $\pp$ invariant mass
versus the decay length in the transverse plane ($L_{xy}$) for events
that satisfy the above selection criteria and have a $\ks$ momentum
between 1.45 and 1.50~GeV/$c$. The cluster of events with mass
consistent with the nominal $\ks$ mass and with a long decay length
indicates a clear $\ks$ signal.

\begin{figure}[htbp]
\centerline{\hbox{
\psfig{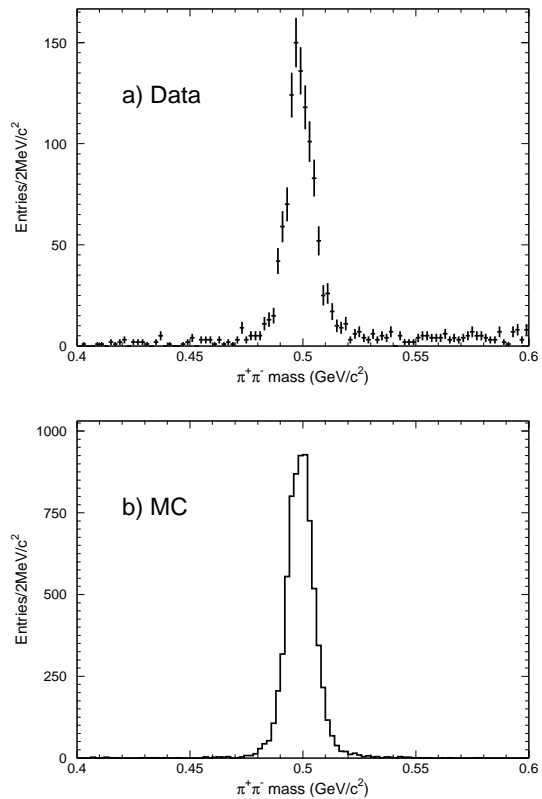}}}
\caption{The $\pp$ invariant mass distributions for
a) data and b) Monte Carlo simulation.}
\label{mks-j}
\end{figure}

Figure~\ref{mks-j} shows the $\pp$ invariant mass distributions 
of both data and Monte Carlo simulation. A fit with a Gaussian 
and a second order polynomial background gives a 
$\ks$ mass of $(499.3\pm 0.2)~\hbox{MeV}/c^2$ and mass resolution
of $(6.5\pm 0.2)~\hbox{MeV}/c^2$ for data, while the 
corresponding numbers are $(499.0\pm 0.1)~\hbox{MeV}/c^2$ 
and $(6.04\pm 0.05)~\hbox{MeV}/c^2$ for Monte Carlo simulation. 
The masses for data and Monte Carlo simulation agree well, 
although both of them deviate from the world average mass 
$(497.672\pm 0.031)~\hbox{MeV}/c^2$~\cite{pdg}. The mass resolution
from Monte Carlo simulation is smaller than that of data.

\begin{figure}[htbp]
\centerline{\hbox{
\psfig{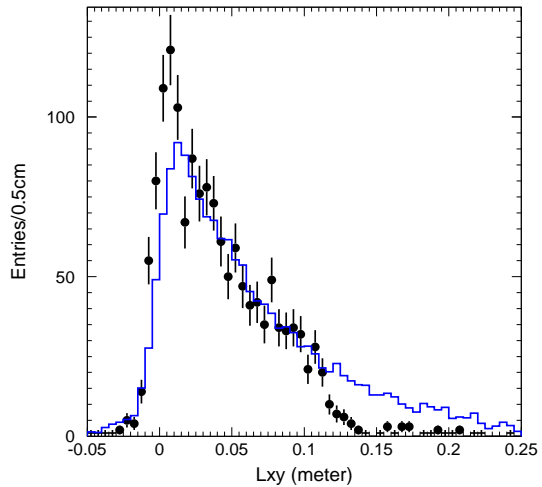}}}
\caption{Comparison of the $\ks$ decay length
distributions between data and Monte Carlo simulation after
normalizing the Monte Carlo data 
to the number of events with $2<L_{xy}<10$~cm. }
\label{lxy-j}
\end{figure}

Figure~\ref{lxy-j} shows the comparison of the $\ks$ decay length
distributions between data and Monte Carlo simulation after
normalizing the Monte Carlo data to the number of events with
$2<L_{xy}<10$~cm.  The discrepancy below 1~cm indicates the still remaining
non-$\ks$ background events in the sample. The difference at $L_{xy}>11$~cm will 
be discussed
later.

\begin{figure}[htbp]
\centerline{\hbox{
\psfig{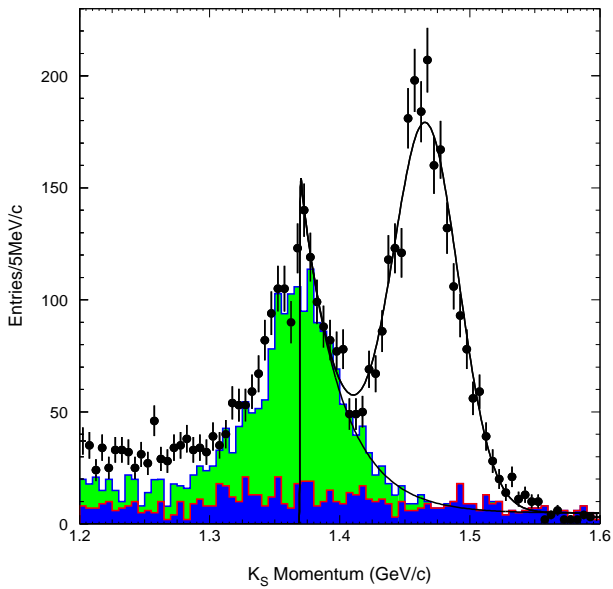}}}
\caption{The $\ks$ momentum distribution for data after the final selection.
Dots with error bars are data, the dark shaded histogram is 
from $\ks$ mass sideband events, and the light shaded histogram 
is the Monte Carlo simulated background. The curve shown in the
plot is from a best fit of the distribution.}
\label{pksfit-j}
\end{figure}

After requiring $L_{xy}>1$~cm and the $\pp$ mass within twice the mass
resolution around the nominal $\ks$ mass and removing the $\gamma$
conversion background (described later), the $\ks$ momentum
distribution is shown in Figure~\ref{pksfit-j}. In the plot, there is
a clear peak around 1.46~GeV/$c$ corresponding to $\jpsito \kskl$
decays, and another peak around 1.37~GeV/$c$ corresponding to
$\jpsito \kskn$.  The background, as estimated from the $\ks$ mass
side bands (three sigma away from the $\ks$ nominal mass on both
sides), can explain the contribution in the high momentum
region, while in the lower momentum region, there are additional backgrounds
due to other channels with $\ks$ production.

\begin{figure}[htbp]
\centerline{\hbox{
\psfig{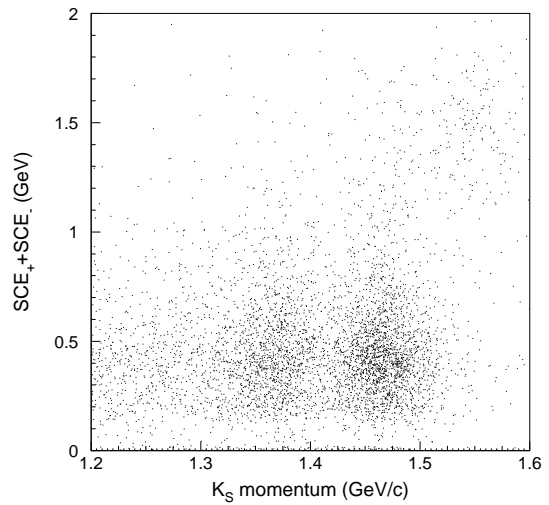}}}
\caption{The total BSC energy of the two charged tracks versus the 
momentum of the $\ks$. The cluster in the upper right corner is due to gamma
conversion background.}
\label{conv-j}
\end{figure}

The secondary vertex requirement and invariant mass cut are very
effective in reducing backgrounds from non-$\ks$ events.  However
since there is no particle identification requirement for the tracks
used, there is contamination from $\EE \ra \gamma \gamma$
where one photon converts into an $\EE$ pair which passes the above selection
criteria.  This background can be seen in Figure~\ref{conv-j}, where the total
BSC energy versus the total momentum of the charged tracks is
shown. The events with high total momentum and large BSC energies in
the upper right corner of the figure are due to this gamma conversion background.

\begin{figure}[htbp]
\centerline{\hbox{
\psfig{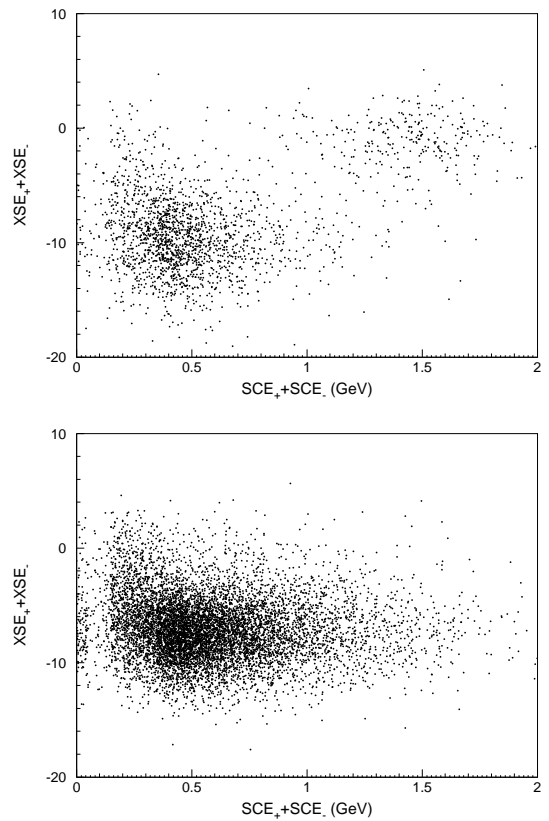}}}
\caption{The scatter plot of the total BSC energy versus the total 
$XSE$ of the two charged tracks with $\ks$ momentum larger than 
1.45~GeV/$c$ for both data (upper) and Monte Carlo simulation (lower).
The cluster in the upper right corner for data is due to gamma conversions.}
\label{scexse-j}
\end{figure}

Figure~\ref{scexse-j} shows a scatter plot of the total BSC energy
versus the total $XSE$ (difference from the expected $dE/dx$
for the electron hypothesis divided by the $dE/dx$ resolution)
of the two charged tracks for events with $\ks$ momentum larger than
1.45~GeV/$c$ for both data and Monte Carlo simulation. It can be seen
that requiring a total BSC energy greater than 1.0~GeV and total $XSE$
greater than $-4$ will select almost all the gamma conversion
background, while the efficiency of this cut for the signal is very
high (about 99.0\% according to Monte Carlo simulation).

\begin{figure}[htbp]
\centerline{\hbox{
\psfig{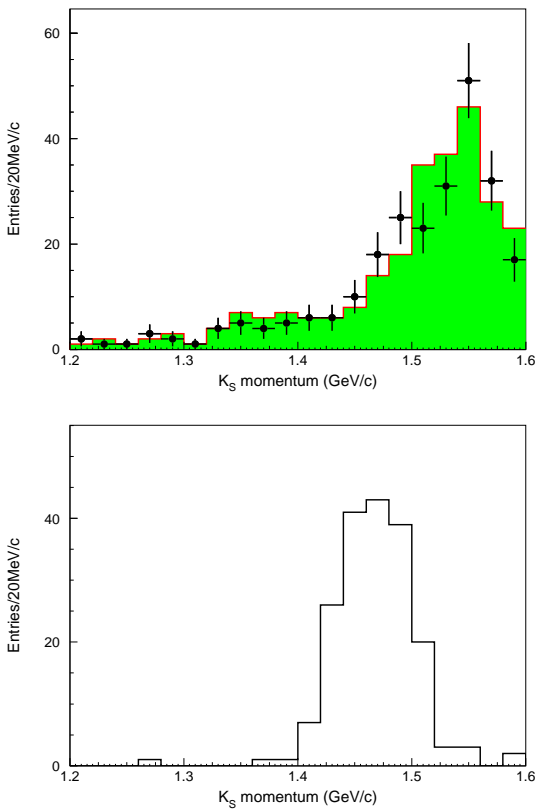}}}
\caption{The $\ks$ momentum distribution of events identified as
gamma conversions. Events in the $\ks$ mass signal region are
shown by dots with error bars and the events in $\ks$ mass
sidebands are shown by the shaded histogram in the upper plot for 
data, and the Monte Carlo simulation of $\jpsito \kskl$ is shown in
the lower plot.}
\label{pks-gconv-j}
\end{figure}

Figure~\ref{pks-gconv-j} shows the distributions of events identified
as gamma conversions for data and Monte Carlo simulated
signal events. There is no indication of signal in the expected
momentum region for $\jpsito \kskl$ events.

The $\gamma$ conversion events can also be removed by cutting on the 
the opening angle between the two charged tracks; a requirement 
that the opening angle be larger than $20^\circ$ removes about the same
fraction of background events with about the same efficiency for
signal events as the cuts used above. This indicates the reliability
of the cuts used for gamma conversion rejection.

\begin{figure}[htbp]
\centerline{\hbox{
\psfig{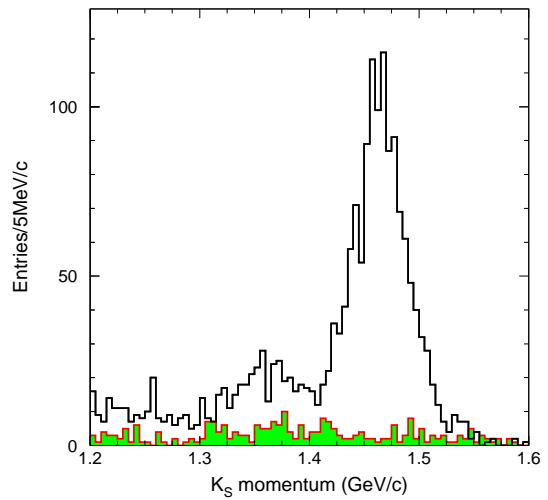}}}
\caption{The $\ks$ momentum distribution for events 
without extra photons (blank histogram).  The $\ks$ sideband
background is shown by the shaded histogram.}
\label{pksngm-j}
\end{figure}

Since there is no photon production in $\kskl$ events, one expects no
photons reconstructed in the candidate events. However the $\kl$ may
decay in the detector, and the decay products or hadronic interactions
of the $\kl$ with the detector material can produce clusters in the
shower counter. As a check, we required the number of photon
candidates in the event to be zero (about 45\% of $\kskl$ events satisfy
this cut
according to Monte Carlo simulation). Figure~\ref{pksngm-j} shows the
$\ks$ momentum distribution after this cut. It is clear that the
background level, including the peak corresponding to $\jpsito
\kskn$, is greatly reduced, while the peak at high momentum is
lowered by about a factor of two as expected from Monte Carlo
simulation.

\section{Backgrounds}

\subsection{Continuum background}

$\kskl$ production via virtual photon annihilation is
forbidden under SU(3) symmetry.
This is checked by applying the same selection criteria
to the data sample taken below the $\jpsi$ peak, at 
$\sqrt{s}=3.0$~GeV. This data was taken during the $\jpsi$
data taking, and the integrated luminosity is measured to
be ${\cal L}=0.7~\hbox{pb}^{-1}$.

\begin{figure}[htbp]
\centerline{\hbox{
\psfig{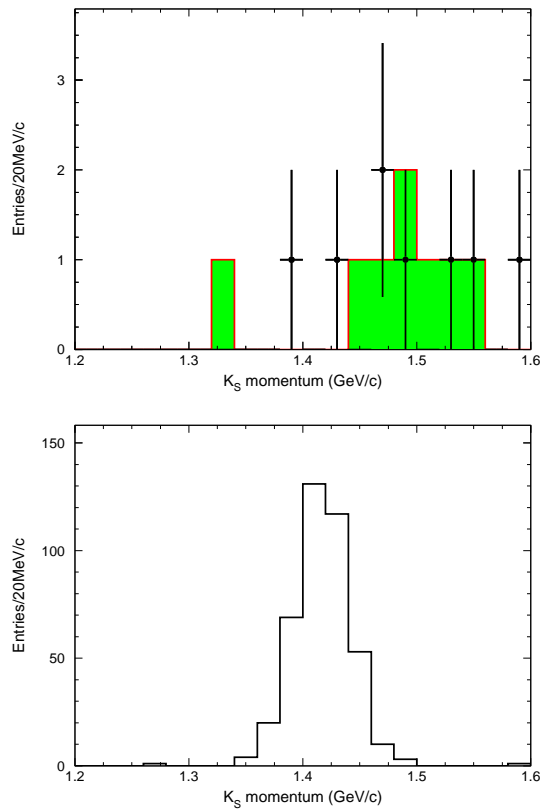}}}
\caption{The $\ks$ momentum distributions for data (upper) and
Monte Carlo simulation (below) at $\sqrt{s}=3.0$~GeV. The dots
with error bars are data, and the shaded histogram in the upper plot
is from the $\ks$ mass sidebands.}
\label{pks-3.0}
\end{figure}

The $\ks$ momentum spectrum of the selected events is shown in
Figure~\ref{pks-3.0}; the events in the signal region agree well with
the expectation from the $\ks$ mass sidebands.
As a conservative estimation, we take all the events with
momentum within two standard deviation from the central
value predicted by the Monte Carlo as signal to set the
upper limit on the production cross section. For the two
observed $\EE\ra \kskl$ candidates, the upper limit on the cross 
section at the 90\% C.~L. is
\[ \sigma<35~\hbox{pb}. \]

The integrated luminosity of the $\jpsi$ data sample is estimated to
be around 17.8~pb$^{-1}$~\cite{fangss}, approximately 25 times as
large as the continuum sample. Since the efficiencies for detecting
$\kskl$ at the $\jpsi$ and at $\sqrt{s}=3.0$~GeV are about the same,
we estimate the continuum contribution of $\kskl$ at the $\jpsi$ to be
at most 50 events, which is small compared to the number of events
observed at the $\jpsi$ (more than 2000).
Since the lack of evidence for $\kskl$ production from the
continuum
agrees well with the SU(3) symmetry prediction, this contribution
is neglected in the following analysis.

\subsection{Backgrounds from inclusive $\jpsi$ decays}

\begin{figure}[htbp]
\centerline{\hbox{
\psfig{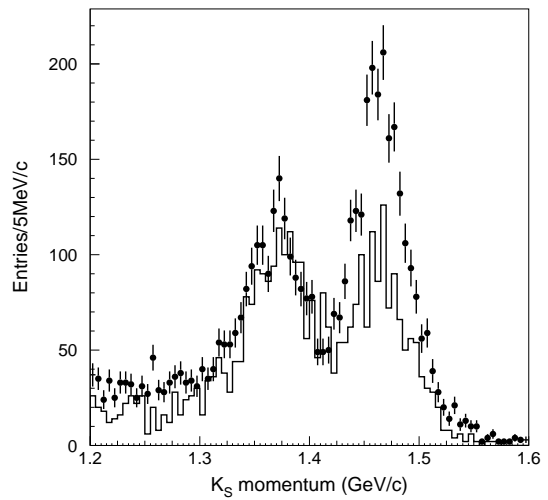}}}
\caption{Comparison of the $\ks$ momentum distributions between data
(dots with error bars) and the inclusive Monte Carlo sample (histogram),
normalized to the total number of $\jpsi$ events. }
\label{pksdtmc-j}
\end{figure}

Figure~\ref{pksdtmc-j} shows the $\ks$ momentum spectrum obtained for the
inclusive Monte Carlo events after all cuts and after normalizing to
the total number of $\jpsi$ events. It can be seen that there
are also two peaks at the expected positions for $\kskn$ and
$\kskl$ as has been observed with data.  The
Monte Carlo simulation reproduces the shape of the peaks, but is
lower than the data.  This indicates the branching ratios used in the
Monte Carlo simulation are too low. The branching fraction of $\jpsito
\kskl$ in the generator is $7.8\times 10^{-5}$, and that of $\jpsito
\kskn$ is $5.08\times 10^{-3}$.

The main background in the intermediate $\ks$ momentum region is due to
$\jpsito \kskn$ where the $K^{*0}$ decays
into $K^0$ and a $\piz$, and one $K^0$ becomes a $\ks$ and the other
becomes a $\kl$. Another potential background is due to $\jpsito
\rho^0\pi^0$, which has a large branching ratio, but this background
is included in the $\ks$ side band events. The background from
$\jpsito \gamma \etac$, with $\etac$ decaying into final states
containing a $\ks$ can also contaminate the signal, but this background
is small because $\etac$ production is two orders of magnitude
lower than $\jpsi$.

Figure~\ref{pksfit-j} shows the $\ks$ momentum distribution 
for the background channels with the input branching fraction
of $\jpsito \kskn$ taken to be $(5.2\pm 0.5)\times 10^{-3}$, obtained from 
a preliminary analysis of the same data sample, together
with the contribution from the $\ks$ mass side band events.
The agreement between the background 
estimation and data is good below the $\kskl$ peak,
indicating the estimation of the background under the $\kskl$ 
peak is reliable. The discrepancy at lower momentum indicates 
backgrounds from other channels (like 
$\jpsito K^{*}(892)^0\overline{K^{*}_0}(1430)^0 + c.c.$,
$\jpsito K^{*}(892)^0\overline{K^{*}_2}(1430)^0 + c.c.$,
etc.), which are not generated in this comparison, are important,
but they do not affect the results in the signal region.

\section{Fit of the momentum spectrum}

The $\ks$ momentum spectrum of the selected events is fitted from 1.37
to 1.60~GeV/$c$ with a Gaussian distribution for the signal, a
constant term for the non-$\ks$ background, and an exponential term
for the background from $\jpsito \kskn$ using an unbinned maximum
likelihood method. The fit results are shown in Figure~\ref{pksfit-j};
the backgrounds from the $\ks$ mass side bands and the $\jpsito \kskn$
background, also shown, agrees well with the fitted background.
The fitted $\ks$ momentum peak is at $(1466.2\pm 0.7)$~MeV/$c$,
which agrees well with the expectation of $1466.3$~MeV/$c$. The fitted
momentum resolution is $(25.2\pm 0.7)$~MeV/$c$, which is in good
agreement with that of the Monte Carlo simulation, $(24.4\pm
0.2)$~MeV/$c$. The fit yields $2155\pm 45$ events, and the efficiency
for detecting $\jpsito \kskl$, with $\ks\to \pp$ is $(38.69\pm
0.23)\%$ from the Monte Carlo simulation.

\section{Efficiency corrections and systematic errors}

The systematic error in the branching ratio measurement comes from
uncertainties in
the efficiencies of the photon energy cut, secondary vertex 
finding, MDC tracking, and the trigger; the branching ratios used; the 
number of $\jpsi$ events; the $\ks$ mass cut; the angular distributions;
etc.

\subsection{Photon energy cut}
\begin{figure}[htbp]
\centerline{\hbox{
\psfig{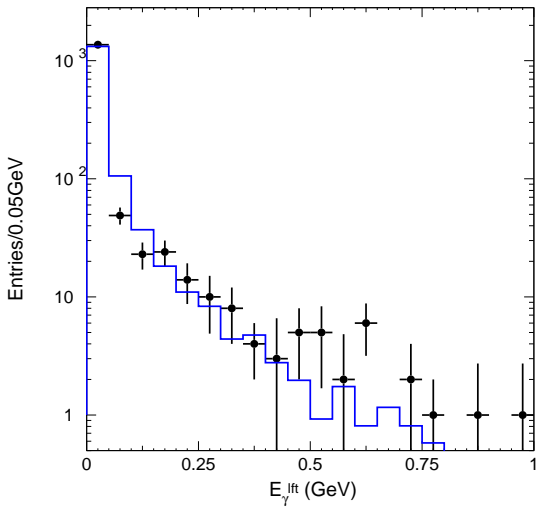}}}
\caption{Comparison of the $E^{lft}_{\gamma}$ distributions for 
$\jpsito \kskl$ events with $\ks$ momentum greater than 1.45~GeV/$c$ 
and less than 1.50~GeV/$c$ between data (dots) and
Monte Carlo simulation (histogram), normalized to the total number of 
events. The contribution of sideband events has been subtracted 
from data.}
\label{eglftj}
\end{figure}

According to the Monte Carlo simulation,
the $E_{\gamma}^{lft}$ cut has an efficiency of 93.6\% for $\jpsito
\kskl$ events, while many backgrounds are removed.  The energy is
produced in signal events by the decays and hadronic interactions of
the $\kl$ with the detector material; the simulation of this effect
depends strongly on the detector simulation software. This is checked
with the $\jpsito \kskl$ signal, requiring the $\ks$ momentum greater
than 1.45~GeV/$c$ and less than 1.50~GeV/$c$. This cut removes almost
all the contamination from $\jpsito \kskn$.  The $E^{lft}_{\gamma}$
distributions of both data and Monte Carlo simulation, which agree
well, are shown in Figure~\ref{eglftj}. The efficiency for data is
found to be $(99.4\pm 0.7)\%$ of that for Monte Carlo simulation.
No
correction to the final efficiency is performed, and 1.5\% is taken as
the systematic error of this cut.

\subsection{Secondary vertex finding}

\begin{figure}[htbp]
\centerline{\hbox{
\psfig{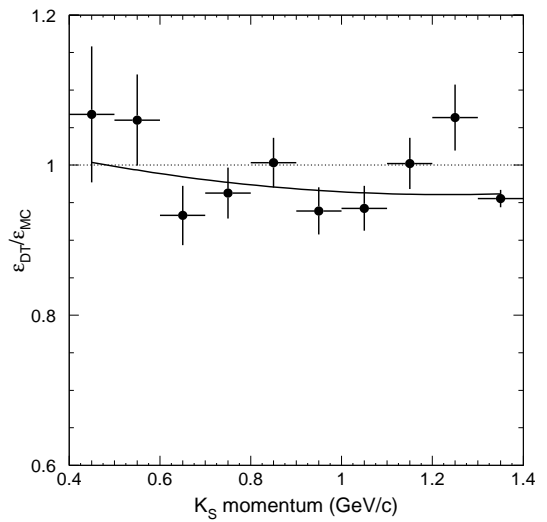}}}
\caption{The ratio of the $\ks$ reconstruction efficiencies of data
and Monte Carlo simulation as a function of the $\ks$ momentum; the
curve shows the best fit to the points using a second order
polynomial.}
\label{rvspks-j}
\end{figure}

The efficiency of the secondary vertex finding algorithm has been checked using $\jpsito
\kskn$ events, where the $\ks$ has a momentum around 1.37~GeV/$c$, and
$\jpsito \kskc$ events, where the $\ks$ momentum is between 0.4 and
1.4~GeV/$c$.  The study shows that the Monte Carlo simulates data (with
$L_{xy}>1.0$~cm) fairly well.  Figure~\ref{rvspks-j} shows the ratio
of the $\ks$ reconstruction efficiencies of data and Monte Carlo
simulation as a function of the $\ks$ momentum. Fitting the points with a
second order polynomial and extrapolating to the $\ks$ momentum for $\jpsito
\kskl$ a correction factor to the efficiency from the
Monte Carlo simulation can be obtained.

\begin{figure}[htbp]
\centerline{\hbox{
\psfig{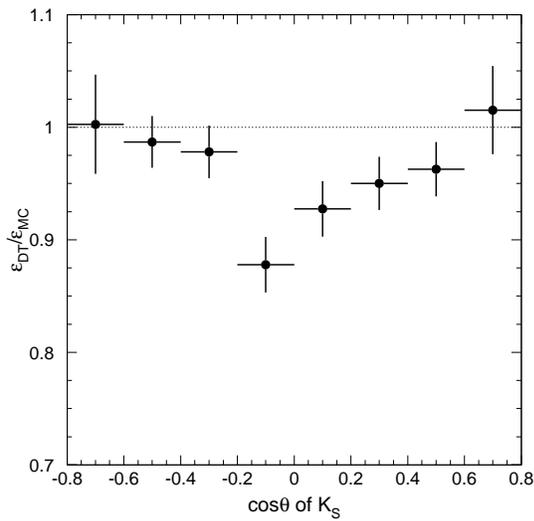}}}
\caption{The ratio between the $\ks$ reconstruction efficiencies
of data and Monte Carlo simulation as a function of the $\ks$
polar angle. }
\label{rvscos-j}
\end{figure}

The polar angle dependence of the $\ks$ reconstruction efficiency has
also been studied with the above sample.  Figure~\ref{rvscos-j} shows
the ratio between the $\ks$ reconstruction efficiencies of data and
Monte Carlo simulation as a function of the cosine of the $\ks$ polar
angle.  Reweighting the efficiency by the expected angular
distribution of the $\ks$ in $\jpsito \kskl$  another correction
factor to the efficiency determined by the Monte
Carlo simulation can be obtained.

Combining the above two effects, a correction of $(96.4\pm 3.1)\%$ to 
the Monte Carlo efficiency is obtained. The error, comes from
the extrapolation and the limited statistics of the samples
used, will be taken as the systematic error of the secondary 
vertex finding.

\subsection{MDC tracking}

The MDC tracking efficiency has been measured using channels like 
$\jpsito \Lambda \overline{\Lambda}$ and
$\pspto \pp \jpsi$, $\jpsito \MM$. It is found that the efficiency of
the Monte Carlo
simulation agrees with that of data within 1-2\% per charged track.
Therefore 4\% will be taken as the systematic error on the tracking efficiency
for the channel of interest.
When the $\pi$ momentum spectrum of the selected $\jpsito \kskl$ events is
compared with that of the Monte Carlo simulation,
good agreement between data and Monte Carlo simulation
is observed in the full momentum range.

\subsection{Trigger efficiency}

The trigger condition which strongly affects the $\kskl$ efficiency is
the requirement of hits in the Vertex Chamber~\cite{zhaodx}, since for
the $\ks$ of interest, the momentum is high (1.466~GeV/$c$ for
$\jpsito \kskl$) and the decay length $\gamma\beta c \tau$, is 7.9~cm,
while the outer radius of the VC is 13.5~cm.  Figure~\ref{lxy-j} shows
the $\ks$ decay length in the $xy$-plane of $\jpsito \kskl$ decays.
There is a sudden drop of efficiency at around
$L_{xy}=11-12$~cm for data, which is not seen with the Monte Carlo sample,
since no trigger simulation is included in the current
version of the Monte Carlo.
Normalizing the Monte Carlo events to the data with $L_{xy}$ between
1~cm and 10~cm and comparing the number of events for all
$L_{xy}$ with the Monte Carlo, yields a correction factor of $(80.1\pm
0.8)\%$ to the Monte Carlo efficiency for $\jpsito \kskl$.

\subsection{Angular distribution}

\begin{figure}[htbp]
\centerline{\hbox{
\psfig{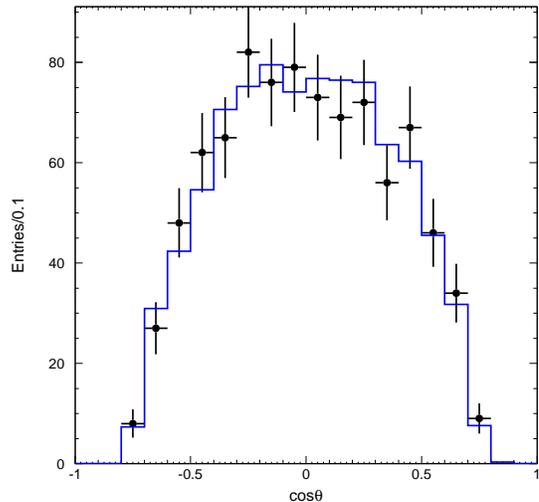}}}
\caption{Cosine of the $\ks$ polar angle of $\kskl$ events
from $\jpsi$ decays. Dots with error bars are data, and the 
histogram is the Monte Carlo simulation.}
\label{cosks-j}
\end{figure}

Figure~\ref{cosks-j} shows the cosine of the $\ks$ polar angle
for $\kskl$ events from $\jpsi$ decays; good agreement
between data and Monte Carlo simulation is observed. This
indicates that the input angular distribution in the Monte Carlo generator 
is correct.

\subsection{Other systematic errors}

The $\ks$ momentum distribution is also fitted between 1.2 and
1.6~GeV/$c$ with a Gaussian smeared Breit-Wigner for the $\kskn$
signal, a Gaussian for the $\kskl$ signal, and a first order polynomial 
for the background. The number of events obtained changes from the
result of the previous fit by 3.3\%.
This is taken as the systematic error due to the uncertainty in the 
background shape.
The number of $\jpsi$ events used in this analysis is taken from
Ref.~\cite{fangss}, and an uncertainty of 4.72\% is used as the 
systematic error.
The systematic error on the branching ratio used, $\BR(\ks \ra \pp)$
is obtained from the Particle Data Group~\cite{pdg} directly.

\subsection{Total systematic error}

Table.~\ref{sys-j} lists the systematic errors from all sources,
as well as the correction factor to the Monte Carlo efficiency.
The total correction factor is 0.772, and the total systematic error 
is 7.2\%.

\begin{table}[htbp]
\caption{Summary of efficiency correction factors and 
         systematic errors.}
\begin{center}
\begin{tabular}{l|c}
\hline\hline
Source  & Corr. factors and syst. errors (\%) \\\hline
Monte Carlo statistics      &  0.6              \\
Photon energy cut           &  1.5                \\  
Secondary vertex finding     &  $96.4\pm 3.1$  \\
MDC tracking                &  4                \\
Trigger efficiency          &  $80.1\pm 0.8$  \\
Background shape            &  3.3            \\
$N_{\jpsi}$                 &  4.7               \\
$\BR(\ks \ra \pp)$          &  0.4               \\
\hline
Total correction factor ($f$) &  77.2 \\\hline
Total systematic error      &   7.2 \\\hline\hline
\end{tabular}
\end{center}
\label{sys-j}
\end{table}

\section{Results and discussion}

The branching ratio of $\jpsito \kskl$ can be calculated from
\[ \BR(\jpsito \kskl)=\left.\frac{n^{obs}/(\eff\cdot f)}
     {N_{\jpsi}\BR(\ks\ra \pp)}
     \right. ~. \]
Using numbers from above (summarized in Table~\ref{br}), one gets
\[ \BR(\jpsito \kskl)
     = (1.82\pm 0.04\pm 0.13)\times 10^{-4} , \]
where the first error is statistical and the second is 
systematic. 
This branching ratio is significantly larger than the 
world average~\cite{pdg} ($(1.08\pm 0.14)\times 10^{-4}$),
which is the combined result of the DMII~\cite{dm2exp}
and MARKIII~\cite{mk3pp} measurements. 

\begin{table}[htbp]
\caption{Numbers used in the branching ratio calculation and final
  branching ratio.}
\begin{center}
\begin{tabular}{l|c}
\hline\hline
quantity      & Value \\\hline
$n^{obs}$     & $2155\pm 45$ \\
$\eff$ (\%)   & $38.69 \pm 0.23$ \\
$f$ (\%)      & $77.2\pm 3.4$ \\
$N_{\jpsi} (10^6)$   &  $57.7\pm 2.7$ \\
$\BR(\ks \ra \pp)$  &  $0.6860\pm 0.0027$ \\\hline
$\BR(\jpsito \kskl) (10^{-4})$ & $1.82\pm 0.04\pm 0.13$  \\
\hline\hline
\end{tabular}
\end{center}
\label{br}
\end{table}

Comparing with the corresponding branching ratio of 
$\pspto \kskl$ ($(5.24\pm 0.47\pm 0.48)\times 10^{-5}$)~\cite{bes2kskl},
and considering the common errors which cancel out in the 
calculation of the ratio between the two branching ratios,
one obtains
      \[ Q_h = \frac{\BR(\pspto \kskl)}{\BR(\jpsito \kskl)}
             = (28.2\pm 3.7)\%. \] This number deviates from the pQCD
predicted ``12\% rule'' by more than 4 standard deviation. Of
particular interest is that $\psp$ decays are enhanced in this channel,
while in almost all other channels where deviations from the
``12\% rule'' are observed, $\psp$ decays are suppressed.

The branching ratio of $\kskl$, along with the branching
ratios of $\jpsito \pp$ ($(1.47\pm 0.23)\times 10^{-4}$)
and $\jpsito \kk$ ($(2.37\pm 0.31)\times 10^{-4}$) from previous 
measurements~\cite{pdg}, can be used to extract the
phase angle difference between the strong and electromagnetic amplitudes
of $\jpsi$ decays into pseudoscalar meson pairs. 
Neglecting the contribution of the continuum in the $\pp$ and
$\kk$ modes, one finds the phase is $\pm(103\pm 7)^\circ$~\cite{wymzprl}.
It should be noted that, since the branching ratio of 
$\jpsito \kskl$ is found significantly larger than previously
measured ones, the branching ratios of $\jpsito \pp$
and $\jpsito \kk$ should also be reexamined.

\section{Summary}

The flavor SU(3) breaking process $\jpsito \kskl$ is measured with
improved precision using BESII data at the $\jpsi$ energy, and the
branching ratio is determined to be 
\( \BR(\jpsito \kskl) = (1.82\pm 0.04\pm 0.13)\times 10^{-4} \), 
which is significantly larger than
previous measurements. Comparing $\BR(\pspto \kskl)$ with
this number, the former is enhanced relative to the pQCD ``12\% rule''
by more than 4$\sigma$. The phase difference between the strong and
electromagnetic decays of the $\jpsi$ into pseudoscalar meson pairs is
determined.

\acknowledgments

   The BES collaboration thanks the staff of BEPC for their 
hard efforts. This work is supported in part by the National 
Natural Science Foundation of China under contracts 
Nos. 19991480, 10225524, 10225525, the Chinese Academy
of Sciences under contract No. KJ 95T-03, the 100 Talents 
Program of CAS under Contract Nos. U-11, U-24, U-25, and 
the Knowledge Innovation Project of CAS under Contract 
Nos. U-602, U-34(IHEP); by the National Natural Science
Foundation of China under Contract No. 10175060 (USTC); 
and by the Department of Energy under Contract 
No. DE-FG03-94ER40833 (U Hawaii).

\end{document}